\documentclass[%
reprint,
superscriptaddress,
 amsmath,amssymb,
 aps,
pra,
floatfix,
toc=flat,
]{revtex4-1}
\usepackage{comment}
\usepackage{graphicx}
\usepackage{hyperref}
\usepackage[usenames,dvipsnames]{xcolor}
\usepackage{braket}
\usepackage{ bbold }
\usepackage{placeins}
\usepackage{lipsum}

\usepackage[utf8x]{inputenc}
\usepackage{color}
\usepackage{bbm} 

\usepackage{amsfonts,amsmath,amssymb,stmaryrd}

\usepackage{subfigure}  
\usepackage{bbm} 
\usepackage{epsfig}
\usepackage{mathrsfs}
\usepackage{verbatim}
\usepackage{centernot}
\usepackage{ulem}
\usepackage{nicefrac}

\DeclareMathOperator{\Tr}{Tr}

\definecolor{blue(pigment)}{rgb}{0.2, 0.2, 0.6}
\definecolor{darkerblue}{rgb}{0.0, 0.0, 0.4}
\definecolor{darkblue}{rgb}{0.0,0.0,0.5}
\definecolor{darkgreen}{rgb}{0.0,0.4,0.0}
\hypersetup{
   colorlinks,
   linkcolor=blue(pigment),
    citecolor=darkgreen,
   urlcolor=darkblue
}
\begin{abstract}
Quantum simulation experiments have started to explore regimes that are not accessible with exact numerical methods. 
In order to probe these systems and enable new physical insights, the need for measurement protocols arises that can bridge the gap to solid state experiments, and at the same time make optimal use of the capabilities of quantum simulation experiments.
Here we propose applying time-dependent photo-emission spectroscopy, an established tool in solid state systems, in cold atom quantum simulators. Concretely, we suggest combining the method with large magnetic field gradients, unattainable in experiments on real materials, to drive Bloch oscillations of spinons, the emergent quasiparticles of spin liquids. We show in exact diagonalization simulations of the one-dimensional $t-J$ model that the spinons start to populate previously unoccupied states in an effective band structure, thus allowing to visualize states invisible in the equilibrium spectrum. The dependence of the spectral function on the time after the pump pulse reveals collective interactions among spinons.
In numerical simulations of small two-dimensional systems, spectral weight appears at the ground state energy at momentum $\mathbf{q} = (\pi,\pi)$, where the equilibrium spectral response is strongly suppressed up to higher energies,  indicating a possible route towards solving the mystery of the Fermi arcs in the cuprate materials. 
\end{abstract}

\begin{document}
\normalem

\title{Visualizing spinon Fermi surfaces with time-dependent spectroscopy}
\author{Alexander Schuckert}
\thanks{These authors contributed equally}
\email[corresponding author emails: ]{alexander.schuckert@tum.de, annabelle.bohrdt@cfa.harvard.edu}
\affiliation{Department of Physics and Institute for Advanced Study, Technical University of Munich, 85748 Garching, Germany}
\affiliation{Munich Center for Quantum Science and Technology (MCQST), 80799 M\"unchen, Germany}

\author{Annabelle Bohrdt}
\thanks{These authors contributed equally}
\email[corresponding author emails: ]{alexander.schuckert@tum.de, annabelle.bohrdt@cfa.harvard.edu}
\affiliation{Department of Physics and Institute for Advanced Study, Technical University of Munich, 85748 Garching, Germany}
\affiliation{Munich Center for Quantum Science and Technology (MCQST), 80799 M\"unchen, Germany}
\address{ITAMP, Harvard-Smithsonian Center for Astrophysics, Cambridge, MA 02138, USA}
\affiliation{Department of Physics, Harvard University, Cambridge, Massachusetts 02138, USA}
\author{Eleanor Crane}
 \affiliation{Department of Electrical Engineering and London Centre for Nanotechnology, University College London, Gower Street, London WC1E 6BT, United Kingdom}
\author{Fabian Grusdt}
\affiliation{Department of Physics and Arnold Sommerfeld Center for Theoretical Physics (ASC),Ludwig-Maximilians-Universit\"at M\"unchen, Theresienstr. 37, D-80333 M\"unchen, Germany}
\affiliation{Munich Center for Quantum Science and Technology (MCQST), 80799 M\"unchen, Germany}
\date{\today}
\maketitle

\textbf{Introduction.--} Just like the back side of the moon is invisible from the earth, certain quantum states may be hidden from standard measurement tools in condensed matter physics. For example, states may be unoccupied at low temperatures or associated with strongly suppressed matrix elements. Akin to the fascination induced by the back side of the moon in popular culture, the back side of the Fermi arcs in the elusive pseudogap phase of the cuprate materials has excited condensed matter physicists for decades. 

The cuprates exhibit superconductivity at unprecedentedly high temperatures~\cite{Bednorz1986}, and fully understanding their phase diagram has become something like a holy grail in the community. 
One particularly intriguing part of this phase diagram is the pseudogap phase. One of its many fascinating properties is the observation of Fermi arcs in angle-resolved photoemission spectroscopy (ARPES)~\cite{Damascelli2003}: around the nodal points $\mathbf{k} = (\pm \pi/2, \pm \pi/2)$, arcs of high spectral weight appear in the spectral function, and in principle could be part of a small Fermi surface \cite{Shen2005,Sachdev2016}. However, these arcs appear to have two endpoints, and the backside of the putative Fermi surface is invisible. An important question is thus whether there exist states on the backside of the Fermi arcs, which are invisible in ARPES measurements. If this is the case, Luttinger's theorem \cite{Luttinger1960} would be violated, as the area enclosed by the putative Fermi surface would be too small, indicating either a thus far unknown broken translational symmetry or topological excitations \cite{Oshikawa2000}. 

Here we propose a scheme to probe unoccupied states~\cite{Kemper2013} in the spectral function of strongly correlated many-body systems, realizable in quantum simulators. It is based on  
\begin{figure*}[!t]
\centering
\includegraphics{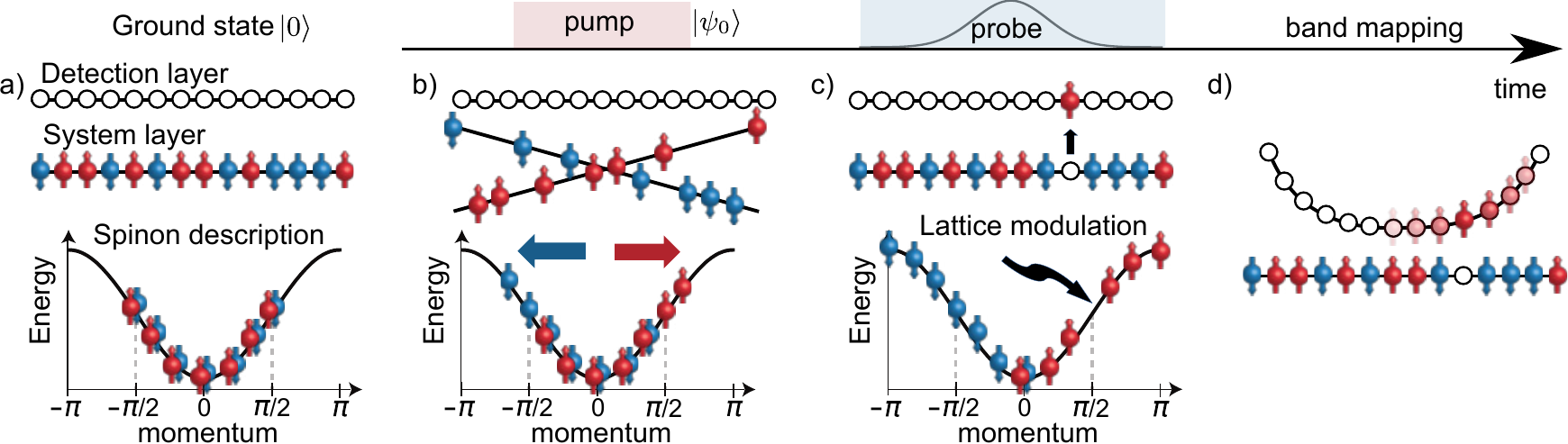}
\caption{\textbf{Time-dependent photoemission-spectroscopy in quantum gases.} a) A half-filled 1D Fermi-Hubbard chain in the ground state, corresponding to a spinon Fermi sea, is prepared in the system layer while the detection layer is empty. b) A strong magnetic field gradient is applied, leading to Bloch oscillations in opposite directions of the two spinon species. c) After the field is switched off, a weak lattice modulation is applied, exciting an atom to the detection layer. d) Finally, the momentum of the excited atom in the detection layer is measured, e.g. in a harmonic potential by a quarter-period oscillation, using time-of-flight or another adiabatic band mapping scheme.}
\label{Fig1}
\end{figure*}
pump-probe spectroscopy, which has recently emerged as a valuable tool in solid state experiments to study non-equilibrium properties of materials~\cite{Schuckert2020,Babadi2017,Giannetti2016,Smallwood2016,Fausti2011,Perfetti2007}.
 Quantum simulators such as ultracold atoms have several advantages: 
 for example, the absence of phonons leads to long coherence times and the Hamiltonian parameters are well known and tunable. In particular, a different toolbox for possible probe pulses is available, such as magnetic field gradients with strengths unattainable in solid state experiments. For these reasons, our scheme is an important complement to existing pump-probe experiments. Our proposal to implement time-dependent ARPES (td-ARPES) to visualize spinon states in cold atoms is within reach of current quantum gas microscopy experiments, which have recently realized all the required building blocks: a magnetic field gradient \cite{nichols2019}, angle-resolved photoemission spectroscopy \cite{Brown2019} and Bloch oscillations~\cite{Dahan1996,Preiss2015,Meinert2017}.

Applying our td-ARPES scheme to the Fermi-Hubbard model, which has been realized with cold atoms in several quantum gas microscopes \cite{Ji2020, Koepsell2020_FL,GuardadoSanchez2020, nichols2019}, directly enables the study of fractionalized excitations: In the one-dimensional (1D) Fermi-Hubbard model, the electron is effectively split into \emph{independent} charge and spin excitations, called chargon and spinon, respectively \cite{Giamarchi,Hilker2017,Vijayan2020}. 
The single-particle spectra of the 1D Fermi-Hubbard and $t-J$ models exhibit a strong asymmetry \cite{Kim1996}, 
which can be associated with the fermionic statistics of spinons and their Fermi sea \cite{Bohrdt2018}. 

In this letter, we demonstrate by numerical simulations that td-ARPES combined with strong external field gradient pulses can shed light on spinon states not occupied in the ground state of the 1D $t-J$ model, up to the highest momentum $k=\pi$, see Fig.~\ref{Fig1}. In two dimensions, we show that a magnetic field gradient pulse along the diagonal direction yields spectral weight at low energies at $k=(\pi,\pi)$. This provides a hint that the missing weight on the back side of the Fermi arcs may be related to a spinon Fermi sea picture \cite{Bohrdt2020_ARPES}. 

\textbf{Measurement Scheme.--} Our protocol combines the equilibrium ARPES protocol~\cite{Bohrdt2018,Brown2019} for quantum simulators with the solid state td-ARPES protocol~\cite{Moritz2010}. First, a system in equilibrium 
is prepared in one layer of an optical lattice. A neighboring layer (``detection layer'') is left empty with a gradient along the transverse direction inhibiting tunneling between the layers due to the energy difference $\Delta$ induced by the gradient, Fig.~\ref{Fig1}a). Subsequently, a non-equilibrium state $\ket{\psi_0}$ is prepared by a quench, such as the application of an external field. Here, we propose to apply a strong magnetic field gradient for a time $t_B$, Fig.~\ref{Fig1}b). Magnetic field gradients have been realized in quantum gas microscopy experiments for example to study spin transport \cite{nichols2019}. 

To measure the time-dependent ARPES spectrum $A_\mathbf{q}(T,\omega)$, we suggest to apply a weak lattice modulation between system and detection layers with frequency $\tilde \omega$ and a Gaussian envelope centered around time $T$ with variance $\Sigma^2$, Fig.~\ref{Fig1}c).
In the weak modulation limit, this allows a single atom of energy $\epsilon_\mathbf{q}$ and spin $\sigma$ to tunnel resonantly into the detection layer if $\tilde \omega=\epsilon_\mathbf{q}+\Delta$. 
Alternatively to the detection layer, a non-interacting third hyperfine state can be used, with the lattice modulation replaced by a radio-frequency (RF) pulse \cite{Brown2019}. This enables the application of our protocol in continuum quantum gases, where RF spectroscopy in equilibrium is routinely performed~\cite{Gupta2003,Greiner2005,Cetina2016}.

Finally, one of the band mapping schemes described in \cite{Bohrdt2018} can be used to measure the momentum of the atom in the detection layer at long times after the pulse has subsided, i.e. $t\gg T+\Sigma$, Fig.~\ref{Fig1}d).
We show in the supplement (SM) \cite{supp} that the momentum space occupation number in the detection layer is proportional to the time-dependent hole spectral function $A_{\mathbf{q}\sigma}(T,\omega)$
at frequency $\omega= \tilde \omega-(\epsilon_\mathbf{q}+\Delta)$ and central time $T$. 
Replacing the empty detection layer by a filled band insulator similarly allows to measure the particle spectral function, see SM~\cite{supp}.

\textbf{Occupying higher momenta.--} 
Our protocol is applicable for fermionic and bosonic systems. Here, we use it to probe unoccupied spinon states in the $t-J$ model, 
\begin{equation}
\hat{\mathcal{H}}_{t-J}= -t \sum_{\langle i,j\rangle,\sigma} \mathcal P\left( \hat c^\dagger_{i,\sigma}\hat c_{j,\sigma} + {\rm h.c.}  \right)\mathcal P+J \sum_{\langle i,j\rangle} \hat{\mathbf{S}}_i \cdot \hat{\mathbf{S}}_{j},
\label{eq:tJ}
\end{equation}
where $\mathcal{P}$ denotes projection on the Hilbert space without double occupancies, $\langle i,j\rangle$ denotes neighboring sites, and $\hat{\mathbf{S}}_j$ are spin-$1/2$ operators.

\begin{figure*}[ht]
\includegraphics{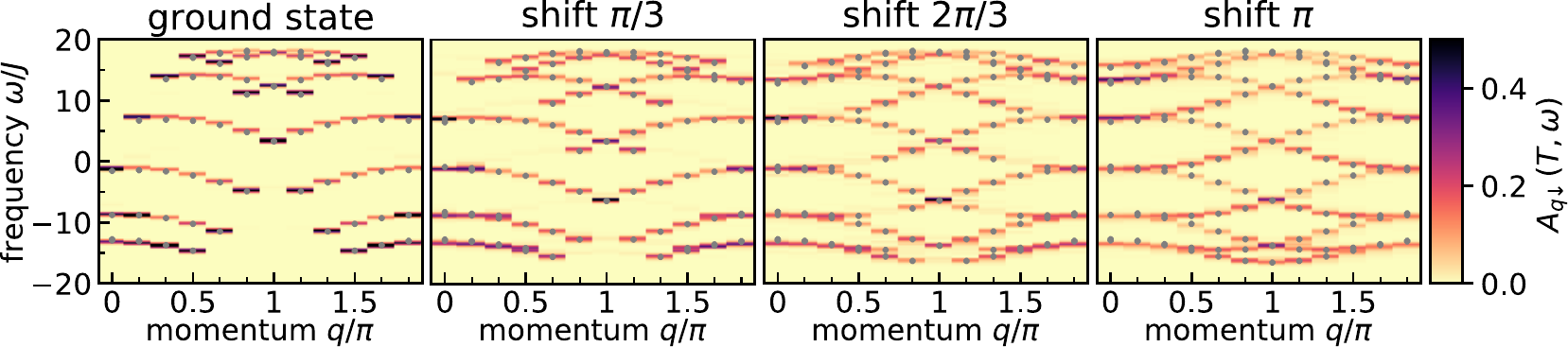}
\caption{\textbf{Occupying spinon states by magnetic field gradient pulses.}
The non-equilibrium one-hole spectral function for a system with $L=12$ sites at $t/J=8$ is shown after a magnetic field gradient was applied. The shift in momentum space is given by the product of magnetic field strength and duration of the gradient, $\theta = Bt_B$, with $t_B=10/J$. Gray dots denote the points where spectral weight is expected 
from our out-of-equilibrium extension of the spectral building principle. The momentum shift due to the magnetic field gradient has been explicitly taken into account in the spinon properties while the chargon stays unaffected. Central times shown are $JT=12.5,15,15$ for shift $\theta = \pi/3, 2\pi/3, \pi$.}
\label{Fig2}
\end{figure*}

The 1D $t-J$ model exhibits spin-charge separation \cite{Giamarchi2003}. This can be made explicit by writing the original fermionic operators as $\hat{c}_{i,\sigma}=\hat{h}_i^\dagger \hat{f}_{i,\sigma}$, where the spin operators $\hat{\mathbf{S}}_i$ are related to the fermionic spinon operators as $\hat{\mathbf{S}}_i = \frac{1}{2} \sum_{\alpha, \beta} \hat{f}_{i,\alpha}^\dagger \boldsymbol{\sigma}_{\alpha,\beta} \hat{f}_{i,\beta}$ and $\hat{h}_i$ denotes the bosonic chargon operator \cite{Baskaran1987,Coleman1984}. 
On a mean-field level, the time dependent spectral function can be approximated as a convolution of spinon and chargon contribution,
\begin{equation}
    A_{q\sigma}(T,\omega)=\sum_{k_h}\int d\nu A^s_{q-k_h\sigma}(T,\omega-\nu)A^c_{k_h}(T,\nu).
    \label{eq:Convolution}
\end{equation}
This representation serves in particular to determine the positions at which spectral weight should appear as it explicitly satisfies momentum and energy conservation. 
Note that due to spin-charge separation, the chargon can be approximated as a free particle with dispersion $\epsilon_h(k)=-2t\cos(k)$ when $t \gg J$ \cite{Ogata1990}.

To gain a better understanding of the spinon contribution to the spectrum, we express the spin part of the Hamiltonian in Eq.~\eqref{eq:tJ} in terms of the spinons \cite{Wen2004}:
\begin{equation}
    \hat{\mathcal{H}}_{J} = -\frac{1}{2} \sum_{\langle i,j\rangle,\alpha} \hat{f}_{i,\alpha}^\dag \hat{f}_{j,\alpha}  \left[ J_\perp  \hat{f}_{j,\bar{\alpha}}^\dag \hat{f}_{i,\bar{\alpha}}+
 J_z   \hat{f}_{j,\alpha}^\dag \hat{f}_{i,\alpha}  \right] 
 \label{eqHfsOriginal}
\end{equation}
where, $\bar{\uparrow}={\downarrow}$ and $\bar{\downarrow}={\uparrow}$. This expression is exact within the subspace satisfying $\sum_\alpha \hat{f}_{i,\alpha}^\dag \hat{f}_{i,\alpha} = 1$ \cite{Wen2004,Bohrdt2018}.

In a mean-field description of the $SU(2)$ invariant model with $J_\perp=J_z$, we replace the operator $\hat{f}_{i,\alpha}^\dagger \hat{f}_{j,\alpha}$ by its ground state expectation value, leading to the formation of a Fermi sea of the spinons $\hat{f}_{i,\sigma}$ \cite{Bohrdt2018,supp}.  
In standard ARPES, fermions can only be removed from formerly occupied states, and thus spectral weight only appears for spinon momenta within the Fermi sea. For an undoped spin chain this corresponds to $k_F = \pm \pi/2$. At momenta $|k|>\pi/2$, states of the many body system exist, but are not occupied and therefore do not yield any weight in the spectral function.

Here, we want to probe these unoccupied spinon states of the 1D $t-J$ model at zero temperature by driving the system out of equilibrium before measuring the time dependent one-hole ARPES spectrum, akin to solid state pump-probe experiments. We do so by applying a magnetic field gradient, described by the Hamiltonian
\begin{equation}
\hat{\mathcal{H}}_B=-B \sum_j j \hat S^z_j,
\label{eq:HB}
\end{equation} 
for a time $t_B$, starting from the $B=0$ ground state.
Numerically, we consider periodic boundary conditions for a cleaner signal and  apply a time-dependent unitary transformation in order to restore translational invariance \cite{supp}.
After time $t_B$, we switch the gradient field off and calculate the td-ARPES spectrum of the resulting non-equilibrium state with exact diagonalization.

In the slave-particle mean-field picture introduced above, the magnetic field gradient exerts an equal but opposite force on the two spinon species, shifting their occupation along the mean-field spinon dispersion. The duration $t_B$ is chosen such that these Bloch oscillations lead to a total shift $\theta = B t_B$, such that after the application of the magnetic field, states with momentum $-\pi/2 \mp\theta/2 \leq k \leq \pi/2 \mp \theta/2$ are occupied by up and down spinons, respectively. The spinons only experience half the total shift since Eq.~\eqref{eq:HB} introduces a coupling of $\mp B/2$ to the density of up/down spinons.

The resulting spinon spectrum then reveals the shifted Fermi seas: spectral weight is obtained for momenta $q$ which are now occupied and were previously empty in the ground state. Within mean-field theory, the  positions of spectral lines can be obtained by inserting the known spinon and chargon dispersions~\cite{Giamarchi, Eder1997,Bannister2000} into Eq.~\eqref{eq:Convolution}.

In Fig.~\ref{Fig2}, we show the numerically obtained spectral function after applying a magnetic field gradient pulse of different strengths, yielding different shifts $\theta/2$. We always remove a spin down particle, thus probing only one of the two spinon Fermi seas. Comparing the numerical results to the spectral building principle \cite{Bannister2000,Eder1997} -- where shifts due to the magnetic field gradient are explicitly taken into account -- yields perfect agreement, providing strong evidence that the slave-particle mean-field theory remains an accurate description beyond the ground state. However, while the mean-field picture predicts a shift of spectral weight in only one direction along the dispersion by $+\theta/2$, we find weight appearing on \emph{both} sides at $\pm \theta /2$. In the following we show that this is due to a time dependence of the spectrum induced by interactions among spinons.  
\\

\begin{figure}[t]
\includegraphics[width=\columnwidth]{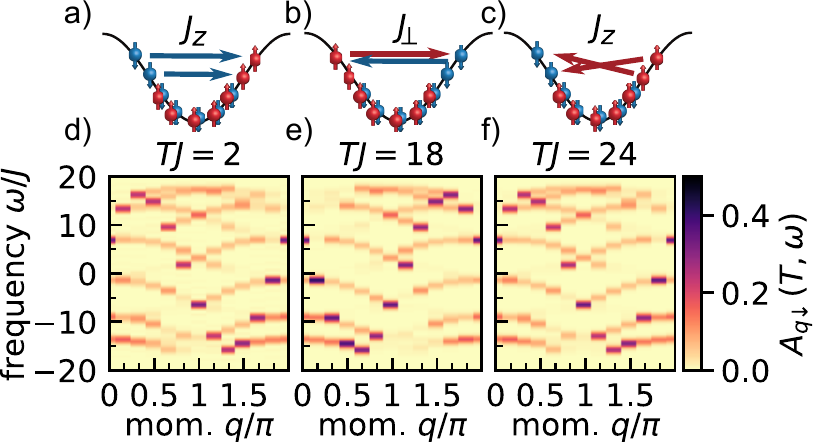}
\caption{\textbf{Coherent oscillations of the spectral weight.} a) Resonant scattering processes leading to a redistribution of spectral weight between the shifted Fermi surfaces. b-d) The spectral function $A_{q\downarrow}(T,\omega)$ is shown as a function of momentum and frequency for three central times $T$ in a system with $L=12$ sites for the case of shift $2\pi/3$ shown in Fig.~\ref{Fig2}. Comparing the spectral function at different central times shows the oscillations of the spectral weight from one side of the visible arcs to the other, corresponding to the Fermi surfaces of the up and down spinon Fermi seas. See also movie in SM~\cite{supp} for full time evolution up to times $TJ\leq 30$.}
\label{Fig3}
\end{figure}

\textbf{Time dependence.--} In the slave-particle mean-field theory, the shifted Fermi sea is still an eigenstate of the Hamiltonian and we thus do not expect to find any dependence on the central time $T$. For sufficiently small magnetic field gradients, the spectral function $A_{q\sigma}(T,\omega)$ indeed does not exhibit a dependence on $T$. However, if the magnetic field gradient is strong, with $B\sim J$, coherent oscillations of the spectral weight emerge, Fig.~\ref{Fig3}. 
 
In particular, the spectral weight in momentum space oscillates between the occupied momenta for up and down spinons, with the spectra for up and down fermions oscillating exactly out of phase, see SM \cite{supp}. However, the position of the spectral lines does not change. 

The Lehmann representation of the spectral function $A_{q\sigma}(T,\omega)$ shows that the only dependence on the time $T$ enters through the eigenstates of the \emph{half-filled} system without a hole \cite{supp}. Therefore, the time-dependence of the spectral function further probes the properties of the spinons in the Heisenberg model. 

Since the time-dependence observed in the spectrum goes beyond a mean-field description, we conclude that interactions between the spinons are relevant, see also \cite{Keselman2020}.
To understand this effect in more detail we examine the quartic slave-particle Hamiltonian in Eq.~\eqref{eqHfsOriginal}. 
First, we consider the $J_z$ term in the Hamiltonian, reading $-J_z/(2L) \sum_{\sigma,qkk'} \cos(q)\hat f^\dagger_{k+q\sigma}\hat f^\dagger_{k'-q\sigma}\hat f_{k'\sigma} \hat f_{k\sigma} $ in momentum space. Enforcing energy conservation on the level of the mean-field dispersion, only processes involving $q=0$, $k'=\pi-k$ or $k'=q+k$ are allowed, examples of which are sketched in Fig.~\ref{Fig3}a),c). This picture is confirmed by switching $J_\perp=0$ after the application of the magnetic field gradient and shifting by $\theta/2=2\pi/L$, which yields perfect sinusoidal oscillations of the up/down Fermi seas, see SM~\cite{supp}. For $J_\perp=J_z$, we find \emph{two} frequency components of the oscillations, which we attribute to the additional resonant coupling of the two Fermi seas by $J_\perp$ processes as sketched in Fig.~\ref{Fig3}b).

\begin{figure}[t]
\includegraphics[width=\columnwidth]{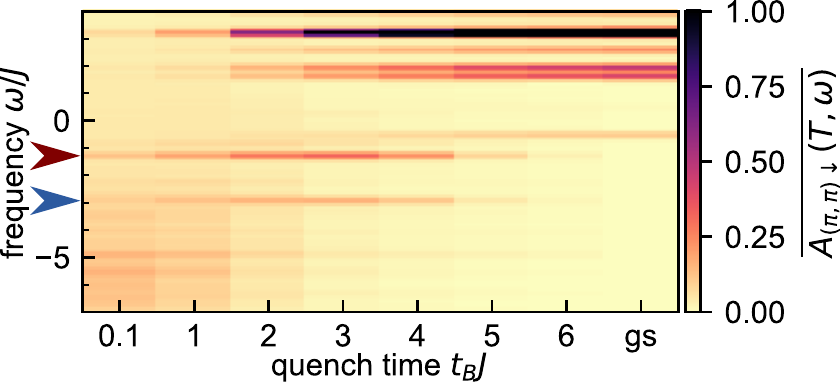}
\caption{\textbf{Time-dependent ARPES in two dimensions.}
The spectral function $A_{\mathbf{q}\downarrow}(T,\omega)$ in the $t-J$ model on a $4\times 4$ torus at $t/J=2$ evaluated at momentum $\mathbf{q}=(\pi,\pi)$ calculated for a magnetic field gradient along the diagonal of strength $B t_B=\pi/2$ for different quench times $t_B$. We average over central times $10\leq JT \leq20$, indicated by the overline. The magnetic polaron ground state (gs) energy at momentum $q=(\pi,\pi)$[$q=(\pi/2,\pi/2)$], $\omega \approx -2.93 J$ [$\omega \approx -1.28 J$] is indicated by a red(blue) arrow.}
\label{Fig2D}
\end{figure}
\textbf{Two-dimensional $t-J$ model.--} 
The same protocol can also be applied in a 2D system. In Fig.~\ref{Fig2D}, we show numerical results for the spectral function $A_{\mathbf{q}\downarrow}(T,\omega)$ in the $t-J$ model on a $4\times 4$ torus at momentum $\mathbf{q}=(\pi,\pi)$. In equilibrium, the spectral weight is strongly suppressed at this momentum at low energies. A possible explanation involves binding of a light chargon to a heavy fermionic spinon from a Fermi sea \cite{Bohrdt2020_ARPES}. This suggests that additional spectral weight should appear at low energies when the magnetic field gradient is switched on and the putative spinon Fermi surface is displaced.

In the time-dependent spectral function, we find that a fast quench along the diagonal with $\theta=\pi$ and quench time $t_BJ\leq 3$ leads to the appearance of a free particle dispersion $-2t(\cos(q_x)+\cos(q_y))$, see SM~\cite{supp}. This indicates that the fast quench yields a ferromagnetic spin environment through which the hole can move freely. 
More interestingly, a quench with $\theta = \pi/2$ at moderate speeds leads to the appearance of spectral weight around the ground state energy of a single hole in the $t-J$ model, indicated by arrows in Fig.~\ref{Fig2D}. We interpret this feature as a signature of a spinon Fermi sea in two dimensions. Finally, for very slow quenches, finite-size gaps prevent any interesting dynamics and the obtained spectrum resembles the ground state result.

\textbf{Summary and Outlook.--} 
We propose a measurement scheme to reveal unoccupied spinon states in the one-hole spectral function. Our numerical results for the 1D $t-J$ model show that a strong magnetic field gradient leads to spinon Bloch oscillations, where the spinons occupy previously empty momentum states. These can then be probed by time-dependent spectroscopy, revealing beyond mean-field interactions among spinons. Extending our results to small 2D systems, we are able to visualize the ground state of the magnetic polaron at momentum $\mathbf{q} = (\pi,\pi)$ in the spectral function, which has no spectral weight in the equilibrium spectrum \cite{Bohrdt2020_ARPES}. 

A promising future direction is to perform similar numerical simulations of the time-dependent spectral function for extended 2D systems. The perhaps most interesting candidates are the 2D Fermi-Hubbard or $t-J$ models at finite doping, where our protocol can help to resolve the long-standing question concerning the existence of (unoccupied) states on the back side of the Fermi arcs. As numerical calculations in this regime are challenging, experimental realizations become essential. Measuring the particle spectral function, which is challenging in solids, could lead to an alternative route to probing unoccupied states.
Apart from the square lattice $t-J$ model discussed here, pump-probe spectroscopy with strong magnetic fields also enables new insights into the properties of spinons in other models and geometries, such as triangular and Kagom\'e lattices or the $J_1-J_2$ model. 
Graphene, simulable in hexagonal lattice quantum simulators, exhibits a d-wave superconducting state due to a Van-Hove singularity, which could be probed by shifting the band structure occupation akin to our protocol~\cite{Black2014}.


\textbf{Acknowledgements.--} We thank Michael Knap for fruitful discussions and comments, and acknowledge useful discussions with Nir Navon, Eugene Demler, Daniel Greif, Waseem Bakr, Christian Gross and Immanuel Bloch.\\

We acknowledge support from the Deutsche Forschungsgemeinschaft (DFG, German Research Foundation) under Germany's Excellence Strategy--EXC-2111--390814868, the Max Planck Gesellschaft (MPG) through the International Max Planck Research School for Quantum Science and Technology (IMPRS-QST), the European Research Council (ERC) under the European Union’s Horizon 2020 research and innovation programme (grant agreement No. 851161), the Technical University of Munich - Institute for Advanced Study, funded by the German Excellence Initiative, the European Union FP7 under grant agreement 291763, from DFG grant No. KN1254/1-1, No. KN1254/2-1, and DFG TRR80 (Project F8), the NSF through a grant for the Institute for Theoretical Atomic, Molecular, and Optical Physics at Harvard University, the Smithsonian Astrophysical Observatory, and the UK Engineering and Physical Sciences Research Council (COMPASSS/ADDRFSS, Grant No.EP/M009564/1).

\bibliography{literature}

\newpage
\newpage
\onecolumngrid
\begin{center}
\textbf{\large Supplemental Materials: Visualizing spinon Fermi surfaces with time-dependent spectroscopy}
\end{center}
\setcounter{equation}{0}
\setcounter{figure}{0}
\setcounter{table}{0}
\setcounter{page}{1}
\makeatletter
\renewcommand{\theequation}{S\arabic{equation}}
\renewcommand{\thefigure}{S\arabic{figure}}

\tableofcontents
\section{Lehmann representation of time-dependent spectral function}

Our protocol probes the time-dependent spectral function \begin{equation}
    A_{\mathbf{q}\sigma}(T,\omega)=\int d\tau e^{i\omega \tau} \Braket{\psi_0| \hat c_{\mathbf{q},\sigma}^\dagger (T+\tau/2) \hat c_{\mathbf{q},\sigma}(T-\tau/2)|\psi_0},
\end{equation}
where $\tau$ is the relative time between creation and annihilation of a Fermion and $\ket{\psi_0}$ is the non-equilibrium initial state.  By introducing the many-body eigenstates with total particle number $N$, $\hat H\ket{n^N}=E_n^N \ket{n^N}$, we can decompose it into time dependent and thermal parts according to
\begin{equation}
A_{\mathbf{q}\sigma}(T,\omega)= A_{\mathbf{q}\sigma}^{\mathrm{th}}(\omega)+ A_{\mathbf{q}\sigma}^{\mathrm{td}}(T,\omega)
\label{eq:nonEqARPES}
\end{equation}
with
\begin{align}
A_{\mathbf{q}\sigma}^{\mathrm{th}}(\omega)&= \sum_n |\braket{\psi_0|n^N}|^2 \sum_l |\braket{l^{N-1}|c_{\mathbf{q}\sigma}|n^N}|^2\delta(\omega-(E_l^{N-1} -E_n^N)),\\
A_{\mathbf{q}\sigma}^{\mathrm{td}}(T,\omega) &= \sum_{n,m\neq n} \braket{\psi_0|n^N}\braket{m^N|\psi_0} e^{i(E_n^N-E_m^N)T}\notag\\&\qquad \times\sum_l \braket{n^N|c_{\mathbf{q}\sigma}^\dagger|l^{N-1}}\braket{l^{N-1}|c_{\mathbf{q}\sigma}|m^N} \delta\left(\omega-\frac{1}{2}\left(E_l^{N-1}-E_m^N\right)-\frac{1}{2}\left(E_l^{N-1}-E_n^N\right)\right).
\end{align}
$A^\mathrm{th}$ is the expected steady state spectral function from the diagonal ensemble. The only time dependence is contained in the phase factors in $A^\mathrm{td}$. Remarkably, they only depend on eigenstates with occupation $N$, i.e. the system before inserting a hole. In our case, this corresponds to the ground state of the half-filled t-J model, i.e. the Heisenberg spin system.
This means that the spinons are solely responsible for a possible time dependence in the spectral function.
\section{Restoring translational invariance}

\subsection{Leaving initial state invariant}
In order to restore translational invariance for our numerics, we apply a time-dependent unitary transformation
\begin{equation}
\hat U(t) = \exp{\left(-i B t \sum_j j \hat S^z_j\right)},
\end{equation}
such that the transformed Hamiltonian $\hat H_{J} + \hat H_B \rightarrow \hat H'$ is given by 
\begin{align}
\hat H'&=\hat U \hat H \hat U^\dagger +i \frac{\partial \hat U}{\partial t}\hat U^\dagger\\
&=J \sum_j \frac{1}{2}\left(e^{-iBt}\hat S^+_{j+1}\hat S^-_{j} +e^{iBt}\hat S^-_{j+1}\hat S^+_{j}\right)+\hat S^z_{j+1}\hat S^z_{j}.
\end{align}
In this interaction picture, time evolution is governed by the Schr\"odinger equation
\begin{equation}
i\frac{d}{dt}\ket{\psi'(t)}=H'(t) \ket{\psi'(t)},
\end{equation}
with initial state $\ket{\psi'(0)}=\hat U(0)\ket{0}=\ket{0}$. Therefore, the time evolved state in the \emph{unrotated} frame is given by
\begin{align}
\ket{\psi_0} &= \hat U^\dagger(t_B) \mathcal{T} e^{-i\int_0^{t_B} H'(t)dt} \ket{0}
\label{eq:trafo_initstate}
\end{align}
After preparation of the state $\ket{\psi_0}$ we calculate the td-ARPES function
$A_{k\sigma}(t_1,t_2)=\braket{\psi_0|\hat c^\dagger_{k\sigma}t_1) \hat c_{k\sigma}(t_2) |\psi_0}$, which contains time evolution under the tJ model. The tJ Hamiltonian is transformed with $U(t_B)$, such that time evolution is generated by
\begin{equation}
\hat H'_{tJ}=\mathcal P\left[ -t \sum_{\langle j,l\rangle,\sigma} e^{-i\frac{B t_B}{2}(j-l)\sigma} \hat c^\dagger_{j,\sigma}\hat c_{l,\sigma}+J \sum_j \frac{1}{2}\left(e^{-iBt_B}\hat S^+_{j+1}\hat S^-_{j} +e^{iBt_B}\hat S^-_{j+1}\hat S^+_{j}\right)+\hat S^z_{j+1}\hat S^z_{j}\right]\mathcal P.
\label{eq:transformedHtJ}
\end{equation} 
Moreover, we need to transform the operators $\hat c^\dagger_{k\sigma}$ via $\hat U(t_B)\hat c^\dagger_{k,\sigma}\hat U^\dagger (t_B)=\hat c^\dagger_{k-\sigma \theta/2,\sigma}$, such that the td-ARPES function in the lab frame can be written as
\begin{equation}
A_{k\sigma}(t_1,t_2)=\braket{\psi_0|\hat U^\dagger(t_B)\hat c^\dagger_{k-\sigma \theta/2,\sigma}(t_1')\hat c_{k-\sigma \theta/2,\sigma}(t_2')\hat U(t_B)|\psi_0},
\end{equation}
where we marked the times with a prime to make explicit that time evolution is done under $\hat H_{tJ}'$. Finally, inserting the time evolved state in Eq.~\eqref{eq:trafo_initstate}, we can rewrite this as
\begin{equation}
A_{k\sigma}(t_1,t_2)=\braket{\overline{ \psi_0}|\hat c^\dagger_{k-\sigma \theta/2,\sigma}(t_1')\hat c_{k-\sigma \theta/2,\sigma}(t_2')|\overline{ \psi_0}}
\end{equation}
where $\ket{\overline{ \psi_0}} = \mathcal{T} e^{-i\int_0^{t_B} H'(t)dt} \ket{0}$ is the ground state evolved under the rotated Hamiltonian. Using this expression, we never need to explicitly act with the unitary $\hat U$ on the state.

\subsection{Eliminating the phase in the hopping term}
For numerical reasons, we want to additionally eliminate the phase in front of the hopping term in Eq.~\eqref{eq:transformedHtJ}. We can do so by shifting the unitary transformation by the corresponding constant phase, such that
\begin{equation}
\hat U(t) = \exp{\left(i B (t_B-t) \sum_j j \hat S^z_j\right)},
\end{equation}
and
\begin{align}
\hat H''(t)&=J \sum_j \frac{1}{2}\left(e^{iB(t_B-t)}\hat S^+_{j+1}\hat S^-_{j} +e^{-iB(t_B-t)}\hat S^-_{j+1}\hat S^+_{j}\right)+\hat S^z_{j+1}\hat S^z_{j}.
\label{eq:shiftedtrans}
\end{align}
Now, the initial state in the rotated frame is given by $\ket{\psi'(0)}=\hat U(0) \ket{0}$, with $\hat U(0)=\exp{\left(i B t_B \sum_j j \hat S^z_j\right)} \neq \mathbb{1}$ such that contrary to before we need to act with a unitary on the initial state. We do this by preparing the ground state of the Hamiltonian $\hat H''(t=0)$ at time $0$.

The other steps performed above go through as before. Notably, $U(t_B)=\mathbb{1}$, such that both the tJ Hamiltonian and the operators $\hat c_k$ for the time evolution in $A_{k\sigma}(t_1,t_2)$ are unchanged.
Hence, the full expression can be written as
\begin{equation}
A_{k\sigma}(t_1,t_2)=\braket{\overline{ \psi_0}|\hat c^\dagger_{k,\sigma}(t_1)\hat c_{k,\sigma}(t_2)|\overline{ \psi_0}}
\label{eq:rotG<}
\end{equation}
where $\ket{\overline{ \psi_0}} = \mathcal{T} e^{-i\int_0^{t_B} H''(t)dt} \ket{0''}$ is the ground state of $H''(t=0)$ evolved under $H''(t)$ in Eq.~\eqref{eq:shiftedtrans}.

To summarize, we prepare the ground state of Hamiltonian $\hat H''(t=0)$, evolve it under the time dependent Hamiltonian $\hat H''(t)$ until time $t=t_B$ and then calculate the spectrum according to Eq.~\eqref{eq:rotG<}, where the time evolution is performed under the (unrotated) t-J model defined in the main text.

\section{Shift of spinon occupation by a magnetic field gradient within mean field theory}

Here we want to show that the magnetic field gradient protocol leads to a shift of the occupation of the spinon dispersion in the Heisenberg chain. We first start with a simple example where a similar effect happens: spinless non-interacting Fermions in an electric field. We then go on to discuss the spinon mean field theory, showing that the shifts observed in the main text can be understood. However, mean field does not yield any time dependence. Finally, we attempt to expand on the mean field theory by a more general Gaussian state to explain the time dependence, however showing that a Gaussian ansatz does also not suffice.

\subsection{Warmup: Noninteracting Fermions in an electric field}
We consider a single fermionic band in 1D in an electric field, described by Hamiltonian
\begin{equation}
\hat H = -J \sum_i (\hat c_i^\dagger \hat c_{i+1} + h.c.) + \Delta \sum_j j \hat n_j.
\end{equation}
By using the same rotating frame as in the previous section, here created by unitary $\hat U(t)=\exp(-i\Delta(t_B-t)\sum_j j \hat n_j)$, the Hamiltonian in momentum space becomes
\begin{equation}
\hat H' (t) = -2 J \sum_k \cos(k-\Delta(t_B-t))\hat c_k^\dagger \hat c_k.
\end{equation}
We then follow the same procedure as in the main text: We prepare the ground state of $\hat H'(t)$ at time $t=0$, given by
\begin{equation}
\ket{0}=\prod_{|k-\Delta t_B|\leq k_F} \hat c_k^\dagger \ket{0},
\end{equation}
where $k_F$ is the Fermi momentum in the absence of a tilt. The time evolution of this state under the time dependent Hamiltonian $\hat H'$ until time $t_B$ can be calculated by noting that $\hat c_k^\dagger (t_B)=\exp(-2Ji \int_0^{t_B} \cos(k-\Delta (t_B-t'))dt') \hat c_k^\dagger$ and hence
\begin{equation}
\overline{\ket{\psi_0}}\equiv\mathcal{T}\exp\left(-i  \int_0^{t_B} \hat H'(t') dt'\right)\ket{0}= \prod_{|k-\Delta t_B|\leq k_F} \exp\left(-\frac{2Ji}{\Delta} \left(\sin(k)-\sin(k-\Delta t_B)\right)\right)c_k^\dagger\ket{0}.
\label{eq:timeevolved}
\end{equation} 
The ARPES spectrum then follows as
\begin{align}
A_k(\omega)&\equiv\int d(t_1-t_2)e^{i\omega(t_1-t_2))} \overline{\bra{\psi_0}}\hat c_k^\dagger(t_1)\hat c_k(t_2)\overline{\ket{\psi_0}} \\
&=2\pi\delta(\omega-2J\cos(k))\overline{\bra{\psi_0}}\hat c_k^\dagger\hat c_k\overline{\ket{\psi_0}}\\
&=2\pi\delta(\omega-2J\cos(k))\Theta(|k-\Delta t_B|-k_F).
\end{align}
This result shows some of the main features of the numerical results for the tJ model in a gradient magnetic field: One can scan along the whole dispersion by changing $\Delta t_B$, occupying states which are not occupied in the ground state without the field. However, in contrast to the tJ model, the result is both independent of central time $t_1+t_2$ and of the strength of the tilt field $\Delta$. 
 Furthermore, the above result can be easily generalized to spinful Fermions in a magnetic field gradient: they just get shifted in opposite directions by $\pm \Delta t_B/2$, where the factor of one-half comes from the fact that $\hat S^z= \frac{1}{2}(\hat n_\uparrow-\hat n_\downarrow)$. In the following, we show how essentially the same result emerges in the Heisenberg model 

\subsection{Constrained Fermion mean field theory with a magnetic field gradient}
We start from the Heisenberg model in a magnetic field gradient
\begin{equation}
H=J\sum_i \frac{1}{2}\left( S_i^+ S_{i+1}^- + S^-_i S^+_{i+1}\right) +\hat S_i^z \hat S_{i+1}^z- B \sum_j j\hat S^z_j
\end{equation}
Introducing constrained Fermion operators by $\hat S^+_i=\hat f_{i\uparrow}^\dagger \hat f_{i\downarrow}, \hat S^z_i=\frac{1}{2}\left(\hat f_{i\uparrow}^\dagger\hat f_{i\uparrow}-\hat f_{i\downarrow}^\dagger\hat f_{i\downarrow}\right)$ with constraint $\sum_\alpha \hat f_{i,\alpha}^\dagger \hat f_{i,\alpha}=1$, we get
\begin{equation}
\hat H= -\frac{J}{2}\sum_{i,\alpha}\hat f_{i,\alpha}^\dagger\hat f_{i+1,\alpha}\left(f_{i+1,\bar\alpha}^\dagger\hat f_{i,\bar\alpha}+f_{i+1,\alpha}^\dagger\hat f_{i,\alpha}\right) - \frac{ B}{2} \sum_\alpha \alpha \sum_j j \hat f_{j\alpha}^\dagger\hat f_{j\alpha}.
\end{equation}
Here, we defined $\alpha=+/-$ for $\uparrow/\downarrow$ and $\bar \uparrow=\downarrow, \bar \downarrow=\uparrow$. Moreover, we neglected all constant terms.

In order to decouple the interactions, we introduce a spin-dependent mean field \begin{equation}
\chi_\alpha=\braket{\hat f_{i,\alpha}^\dagger \hat f_{i+1,\alpha}}
\end{equation}
and neglect all terms quadratic in the fluctuations around this mean field, arriving at
\begin{equation}
\hat H= -\frac{JL}{2}\sum_\alpha \chi_\alpha (\chi_{\bar \alpha}^\dagger+ \chi_\alpha^\dagger )-\frac{J}{2}\sum_{i,\alpha}\left(\left( \chi_\alpha+\chi_{\bar\alpha} \right)\hat f_{i+1,\alpha}^\dagger \hat f_{i,\alpha} + h.c.\right) - \frac{ B}{2} \sum_\alpha \alpha \sum_j j \hat f_{j\alpha}^\dagger\hat f_{j\alpha}.
\end{equation}
In order to follow the protocol discussed in the main text, we move into the rotating frame of the magnetic field gradient with the unitary 
\begin{equation}
\hat U= \exp\left(i\frac{ B}{2}(t_B-t) \sum_\alpha \alpha \sum_j j \hat f_{j,\alpha}^\dagger\hat f_{j,\alpha}\right), 
\end{equation}
such that
\begin{equation}
\hat f_{j\alpha}\rightarrow Uf_{j\alpha}U^\dagger = \exp\left(-i\frac{\alpha B j}{2}(t_B-t) \right) \hat f_{j,\alpha}.
\end{equation}
Defining $f_j=\frac{1}{\sqrt{L}}\sum_k e^{-ikj} f_k$, and taking the infinite system size limit $(1/L)\sum_k = (1/2\pi) \int dk$ the Hamiltonian density becomes
\begin{equation}
\hat H(t)= -\frac{JL}{2}\sum_\alpha \chi_\alpha(t) (\chi_{\bar \alpha}^\dagger(t)+ \chi_\alpha^\dagger(t) )+\int_{-\pi}^\pi dk \sum_{\alpha}\epsilon_\alpha(k,t) \hat f_{k\alpha}^\dagger \hat f_{k\alpha}
\end{equation}
with dispersion
\begin{equation}
\epsilon_\alpha(k,t) = -\frac{J}{2}\left(\left( \chi_\alpha+\chi_{\bar\alpha}\right) e^{i(k+\frac{\alpha B}{2}(t_B-t))}+c.c.\right).
\end{equation}
Moreover, using that in the ground state $\braket{\hat f^\dagger_k \hat f_{k'}}\sim \delta_{kk'}$, the mean field transforms to
\begin{equation}
\chi_\alpha(t)=\frac{1}{2\pi}\int_{-\pi}^\pi dk e^{-i(k+\frac{\alpha B}{2}(t_B-t))} \braket{\hat f^\dagger_{k,\alpha} \hat f_{k,\alpha}} .\label{eq:chia}
\end{equation}
In order to find the ground state solution for $t=0$, we need to  minimize $\braket{\hat H}$ self-consistently under the constraint \eqref{eq:chia}. To do so, we set the phase of $\chi_\alpha+\chi_{\bar\alpha}$ to zero (without loss of generality as the ground state is degenerate with respect to this phase) such that
\begin{equation}
\epsilon_\alpha(k,t=0) = -J\left(|\chi_\alpha+\chi_{\bar\alpha}| \cos\left(k+\frac{\alpha B}{2} t_B\right)\right).
\end{equation}
The ground state is hence given by a Fermi sea with the Fermi momenta given by $|k_F+\alpha \frac{ B}{2}t_B|=\frac{\pi}{2}$. Inserting this into $\eqref{eq:chia}$, we find 
\begin{align}
\chi_\alpha(t=0)&=\frac{1}{2\pi}\int_{(-\pi-\alpha B t_B)/2}^{(\pi-\alpha B t_B)/2} dk\, e^{-i(k+\frac{\alpha B}{2}t_B)} \\
&=\frac{1}{\pi}
\end{align}
\emph{independent} of $\alpha$ and we get our final result for the spinon dispersion
\begin{equation}
\epsilon_\alpha(k,t=0) = -\frac{2J}{\pi} \cos\left(k+\frac{\alpha B}{2} t_B\right).
\end{equation}

Having found the ground state at $t=0$, we can now proceed with time evolving the state until $t=t_B$ to then calculate the spinon spectral function as in the previous example for free Fermions, yielding
\begin{align}
A^s_{k,\alpha}(\omega)&\equiv\int d(t_1-t_2)e^{i\omega(t_1-t_2))} \overline{\bra{\psi_0}}\hat f_k^\dagger(t_1)\hat f_k(t_2)\overline{\ket{\psi_0}} \\
&=2\pi\delta\left(\omega+\frac{2J}{\pi}\cos(k)\right) \Theta(|k+\alpha B t_B/2|-\frac{\pi}{2}).
\end{align}
This result reproduces the numerical observation that the spin up/down components are shifted in opposite directions by $B t_B/2$. However, it does not contain any time dependence. Hence, we can attribute these effects to beyond-mean field interaction effects of the spinons. In the following, we show that even a more general Gaussian state ansatz can not explain the time dependence.
 
\subsection{Gaussian state}

\subsubsection*{Spinon Hamiltonian in momentum space}

We transform the spinon Hamiltonian 
\begin{equation}
    \hat H= -\frac{J_\perp}{2}\sum_{i\sigma} \hat f^\dagger_{i\sigma}\hat f^\dagger_{i+1\bar \sigma}\hat f_{i\bar\sigma}\hat f_{i+1\sigma}-\frac{J_z}{2}\sum_{i\sigma} \hat f^\dagger_{i\sigma}\hat f^\dagger_{i+1 \sigma}\hat f_{i\sigma}\hat f_{i+1\sigma} 
\end{equation}

into momentum space by defining
\begin{equation}
    \hat f_{k\sigma}=\frac{1}{\sqrt{L}}\sum_j e^{ikj} \hat f_{j\sigma},
\end{equation}
arriving at
\begin{equation}
    \hat H = -\frac{1}{2L} \sum_{kk'q,\sigma} \cos(q) \left[J_\perp \hat f^\dagger_{k+q\sigma}\hat f^\dagger_{k'-q\bar\sigma}\hat f_{k\bar\sigma}\hat f_{k'\sigma} +J_z \hat f^\dagger_{k+q\sigma}\hat f^\dagger_{k'-q\sigma}\hat f_{k\sigma}\hat f_{k'\sigma}\right].\label{eq:Ham}
\end{equation}

\subsubsection*{Gaussian variational ansatz}

In an attempt to explain the oscillations observed in the main text, we introduce a variational state
\begin{equation}
    \ket{\Psi(t)}= \prod_{k_-\leq k\leq k_+} \left( \alpha_k \ket{k,\uparrow}+\beta_k \ket{-k,\uparrow}\right)\prod_{k_-\leq k\leq k_+} \left( \beta_k \ket{k,\downarrow}+\alpha_k \ket{-k,\downarrow}\right)\prod_{0<k<k_-} \ket{k,-k,\uparrow}\ket{k,-k,\downarrow},
\end{equation}
where in this definition $\pi>k\geq 0$ (in the following $\pi>k\geq -\pi$) and $\ket{k,\sigma}=\hat f_{k\sigma}^\dagger\ket{0}$, $\ket{k,-k,\sigma}=\hat f_{k\sigma}^\dagger\hat f_{-k\sigma}^\dagger\ket{0}$. From the normalisation of the state it follows that $|\alpha_k|^2+|\beta_k|^2=1$. All parameters are time dependent, i.e. $\alpha_k=\alpha_k(t)$, $\beta_k=\beta_k(t)$.

Due to the product nature of the state, a generalized Wick's theorem holds in which all higher order correlation functions can be decomposed in two-point correlation functions, which we calculate in the following.
\subsubsection*{Two point correlation functions} 
 The pairing and cross-spin correlation functions vanish,
\begin{equation}
\braket{\hat f^\dagger_{k\sigma}\hat f^\dagger_{k'\sigma}}=\braket{\hat f_{k\sigma}\hat f_{k'\sigma}}=\braket{\hat f^\dagger_{k\sigma}\hat f_{k'\bar\sigma}}=0.
\end{equation}
Moreover, the only remaining combination is nonzero only if $k=k'$ or $k=-k'$, i.e.
\begin{equation}
\braket{\hat f^\dagger_{k\sigma}\hat f_{k'\sigma}}= \delta_{k,k'} \braket{\hat f^\dagger_{k\sigma}\hat f_{k\sigma}}+\delta_{k,-k'} \braket{\hat f^\dagger_{k\sigma}\hat f_{-k\sigma}}.
\end{equation}
Finally, the two terms on the right hand side can be evaluated in the variational state, yielding
\begin{align}
\braket{\hat f^\dagger_{k\uparrow}\hat f_{k\uparrow}}=\begin{cases} 1 & \text{for } |k| < k_- \\ |\beta_k|^2 &\text{for } -k_+\leq k \leq -k_- \\ |\alpha_k|^2 &\text{for } k_-\leq k \leq k_+ \\0 & \text{for } |k| > k_+\end{cases}, \qquad \braket{\hat f^\dagger_{k\downarrow}\hat f_{k\downarrow}}=\begin{cases} 1 & \text{for } |k| < k_- \\ |\alpha_k|^2 &\text{for } -k_+\leq k \leq -k_- \\ |\beta_k|^2 &\text{for } k_-\leq k \leq k_+ \\0 & \text{for } |k| > k_+\end{cases},\\
\braket{\hat f^\dagger_{k\uparrow}\hat f_{-k\uparrow}}=\begin{cases} 0 & \text{for } |k| < k_- \\ \alpha_k\beta_k^* &\text{for } -k_+\leq k \leq -k_- \\ \alpha_k^*\beta_k &\text{for } k_-\leq k \leq k_+ \\0 & \text{for } |k| > k_+\end{cases}, \qquad \braket{\hat f^\dagger_{k\downarrow}\hat f_{-k\downarrow}}=\begin{cases} 0 & \text{for } |k| < k_- \\ \alpha_k^*\beta_k &\text{for } -k_+\leq k \leq -k_- \\ \alpha_k\beta_k^* &\text{for } k_-\leq k \leq k_+ \\0 & \text{for } |k| > k_+\end{cases}.
\end{align}

From the above, we note the following symmetries between the spin correlation functions:
\begin{align}
\braket{\hat f^\dagger_{k\uparrow}\hat f_{k\uparrow}} &= \braket{\hat f^\dagger_{-k\downarrow}\hat f_{-k\downarrow}},\\
\braket{\hat f^\dagger_{k\uparrow}\hat f_{-k\uparrow}} &= \braket{\hat f^\dagger_{-k\downarrow}\hat f_{k\downarrow}}= \braket{\hat f^\dagger_{k\downarrow}\hat f_{-k\downarrow}}^*,
\end{align}
such that we can follow that there are in fact only two independent correlation functions,
\begin{align}
    n_k&=\braket{\hat n_k} \equiv \braket{\hat f^\dagger_{k\uparrow}\hat f_{k\uparrow}}\\
    m_k&=\braket{\hat m_k} \equiv \braket{\hat f^\dagger_{k\uparrow}\hat f_{-k\uparrow}}
\end{align}

\subsubsection*{Equations of motion}

We can now derive the equations of motion for the correlators $n_k$ and $m_k$ from Heisenberg equations of motion for $\hat n_k$, $\hat m_k$,

\begin{align}
    -i\partial_t \hat n_k &= [\hat H, \hat n_k]\\
    &= -\frac{1}{L} \sum_{k',q} \cos(q) \big( J_\perp \hat f^\dagger_{k'+q\uparrow}\hat f^\dagger_{k-q\downarrow}\hat f_{k'\downarrow}\hat f_{k\uparrow}-J_\perp \hat f^\dagger_{k'+q\downarrow}\hat f^\dagger_{k\uparrow}\hat f_{k'\uparrow}\hat f_{k+q\downarrow}\notag\\
    &\qquad\qquad\qquad\,\,\,\, +  J_z \hat f^\dagger_{k'+q\uparrow}\hat f^\dagger_{k-q\uparrow}\hat f_{k'\uparrow}\hat f_{k\uparrow}-J_z \hat f^\dagger_{k'+q\uparrow}\hat f^\dagger_{k\uparrow}\hat f_{k'\uparrow}\hat f_{k+q\uparrow}\big),\\
        -i\partial_t \hat m_k &= [\hat H, \hat m_k]\\
    &= -\frac{1}{L} \sum_{k',q} \cos(q) \big( J_\perp \hat f^\dagger_{k'+q\uparrow}\hat f^\dagger_{k-q\downarrow}\hat f_{k'\downarrow}\hat f_{-k\uparrow}-J_\perp \hat f^\dagger_{k'+q\downarrow}\hat f^\dagger_{k\uparrow}\hat f_{k'\uparrow}\hat f_{-k+q\downarrow}\notag\\
    &\qquad\qquad\qquad\,\,\,\, +  J_z \hat f^\dagger_{k'+q\uparrow}\hat f^\dagger_{k-q\uparrow}\hat f_{k'\uparrow}\hat f_{-k\uparrow}-J_z \hat f^\dagger_{k'+q\uparrow}\hat f^\dagger_{k\uparrow}\hat f_{k'\uparrow}\hat f_{-k+q\uparrow}\big).
\end{align}
Evaluating these equations of motion with respect to the variational state while employing Wicks theorem, i.e. $\braket{f^\dagger_{k_1\sigma_1}f^\dagger_{k_2\sigma_2}f_{k_3\sigma_3}f_{k_4\sigma_4}}\approx \braket{f^\dagger_{k_1\sigma_1}f_{k_4\sigma_4}}\braket{f^\dagger_{k_2\sigma_2}f_{k_3\sigma_3}}-\braket{f^\dagger_{k_1\sigma_1}f_{k_3\sigma_3}}\braket{f^\dagger_{k_2\sigma_2}f_{k_4\sigma_4}}$, we get
\begin{equation}
    \partial_t \braket{\hat n_k}=0,
\end{equation}
showing that this variational state does not suffice to explain the time dependence seen in the main text.

\section{Additional data about coherent oscillations}

\subsection{Oscillations without $J_\perp$}

\begin{figure}
    \centering
    \includegraphics{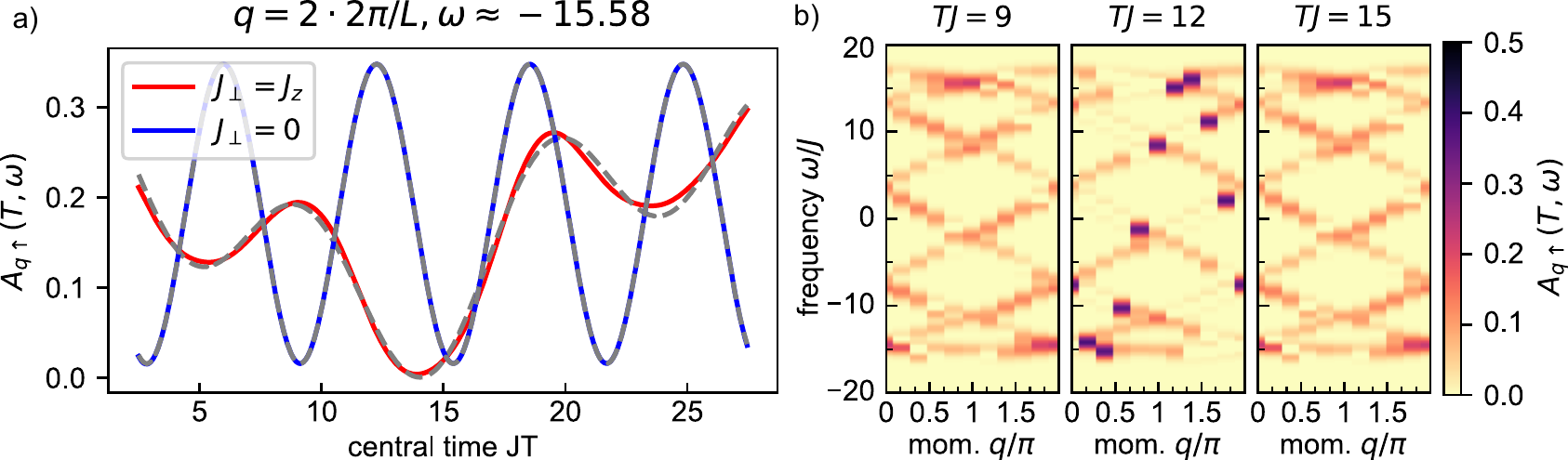}
    \caption{\textbf{Comparing coherent oscillations of spectrum with and without $J_\perp$.--} a) Central time dependence of the spectral line with lowest frequency for a single momentum. Grey dashed lines indicate a fit with a single sine function with frequency $\approx 1J_z$ for $J_\perp=0$ and a sum of two sine functions with frequencies $0.25J$ and $0.64J$ b) Snapshots of the spectrum at the maxima and minima of the oscillations for $J_\perp=0$. $L=10$, $t=8J$. Total shift $\Theta=2\cdot 2\pi/L$.}
    \label{fig:Jp0}
\end{figure}

For a shift by a single momentum point, $\Delta k = 2 \pi/L$, there are only very few processes allowed by the spinon Hamiltonian written in the main text when imposing both momentum and energy conservation. This becomes particularly extreme in the case when $J_\perp=0$. Then, only one state has the same energy as the shifted Fermi sea and is also connected to the shifted Fermi sea by a momentum-conserving process: its ``conjugate'' partner, i.e. the Fermi sea shifted in the opposite direction in momentum space. Within this picture, the spectrum is expected to perform Rabi oscillations between the shifted Fermi sea and its conjugate partner. In Fig.~\ref{fig:Jp0} we show the time evolution of the spectrum when switching $J_\perp=0$ after $t_B$. The Fermi sea then performs perfect sinusoidal oscillations with frequency $J_z$, as we show by looking at the spectral weight of a single line (all others perform the same oscillations). This supports the picture discussed above. However, when looking at snapshots of the whole spectrum, we see that the conjugate state is never fully reached, which may be attributed to a detuning between the shifted Fermi sea and its conjugate introduced by non-energy-conserving processes not included in this picture. 

Contrarily, when keeping $J_\perp=J_z$, we find an oscillation with a superposition of two sine functions, which we attribute to the coupling of the two oscillating Fermi seas by $J_\perp$.

\subsection{Out of phase oscillations of $\uparrow$ and $\downarrow$}
\begin{figure}
    \centering
    \includegraphics{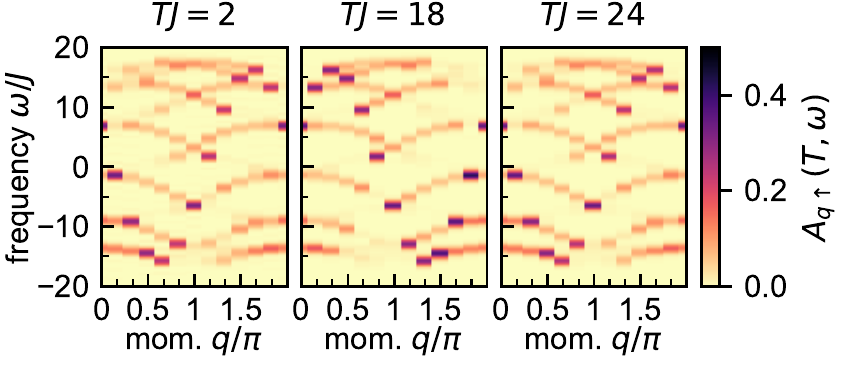}
    \caption{\textbf{Oscillations of $\uparrow$ spectrum.--} Here, we use the same parameters as in Fig.3 of the main text.}
    \label{fig:uposci}
\end{figure}

In Fig.~\ref{fig:uposci} we show the spectrum of $\uparrow$ Fermions at the same times and parameters shown in Fig.3 of the main text. We find perfectly out of phase oscillation compared to the $\downarrow$ Fermions, showing the that the magnetic field indeed induces an opposite shift for both species as predicted in the mean field theory.
\section{Additional data about 2D}

\subsection{Full frequency range for $(\pi/2,\pi/2)$ shift}

\begin{figure}
    \centering
    \includegraphics{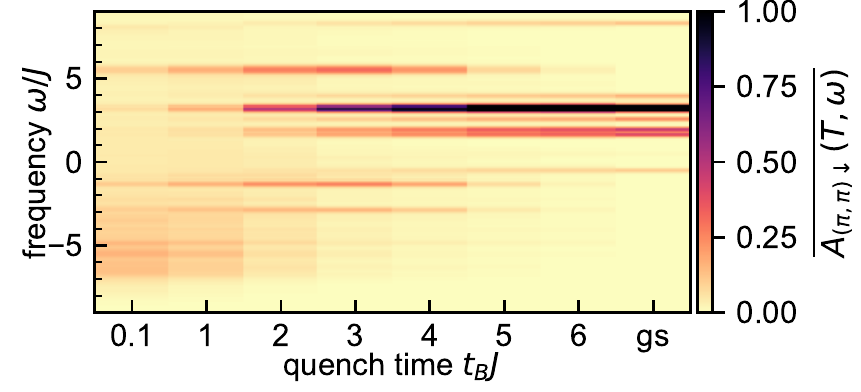}
    \caption{\textbf{Full frequency range of Fig.4 of main text.}}
    \label{fig:4full}
\end{figure}

In Fig.4 of the main text we only showed part of the frequency range in order to emphasize the low-energy feature we attribute to the magnetic polaron ground state energy. In Fig.~\ref{fig:4full} we show the full frequency range, in particular showing a high energy feature around $\omega\approx 5$ appearing for the same values of $t_B$ at which the above mentioned low-energy feature appears.

\subsection{Induced Ferromagnet for $(\pi,\pi)$ shift}

\begin{figure}
    \centering
    \includegraphics{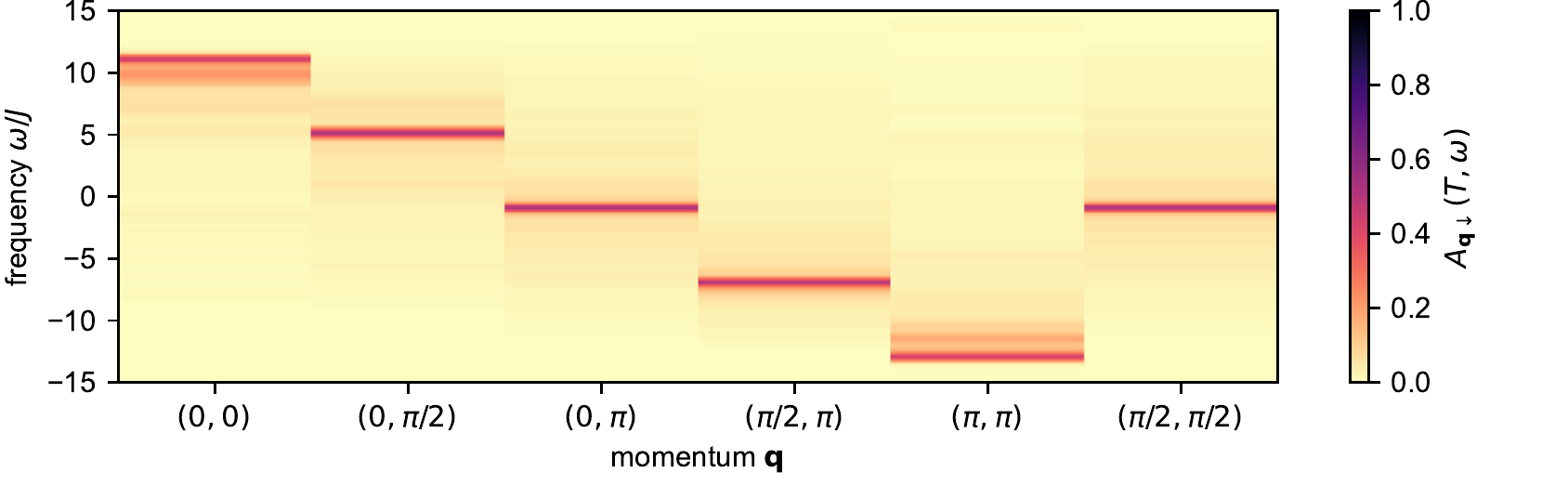}
    \caption{\textbf{Emergence of free particle spectrum for $(\pi,\pi)$ shift in 2D.--} Here, we use $t=3J$ as well as $t_B=1/J$.}
    \label{fig:2Dfree}
\end{figure}

When employing a very strong magnetic field gradient with pulse duration $t_B<1/J$ and total shift is $\Theta=(\pi,\pi)$ , we find the spectrum to be identical to the one of a free particle hopping with dispersion $-2t(\cos(q_x)+\cos(q_y))$, see Fig.~\ref{fig:2Dfree}. We can explain this by considering the strong magnetic field gradient limit, in which we can neglect the Heisenberg Hamiltonian during the dynamics and the system only evolves due to the gradient field. The initial state of the dynamics is given by the ground state of the Heisenberg model in 2D, which shows antiferromagnetic correlations. Now, when a $(\pi,\pi)$ pulse is applied, these correlations get rotated around the z-axis such that in the xy plane there are now ferromagnetic correlations. In a ferromagnet, an injected hole can move freely as there is no energy cost related to the reshuffling of spins associated to the movement of the hole. Hence, the spectrum becomes identical to the one of a free particle.

\section{Time resolved spectroscopy in cold atom experiments}
We show how to measure the time resolved ARPES and ARIPES (inverse ARPES) in cold atom experiments. We do so by combining the ARPES protocol presented in Bohrdt et al.~\cite{Bohrdt2018} to arbitrary initial states,  closely following the theoretical description of time resolved ARPES in the solid state by Freericks et al.~\cite{Freericks2009}. We then generalize the protocols to ARIPES.
\subsection{Linear and quadratic response}
Consider a system with many-body Hamiltonian $\hat H_0$ and probe Hamiltonian $\hat V$ (e.g. containing a lattice modulation or RF pulse), such that the total Hamiltonian is given by $\hat H=\hat H_0 + \hat V$. We work in an interaction picture with respect to $\hat H_0$ such that operators are evolved according to $\hat A_I(t)=\hat U_0^\dagger(t) \hat A \hat U_0(t)$, with $\hat U_0(t)=\mathcal{T}\exp(-i\int dt'\hat H_0(t'))$. The initial density matrix $\hat \rho_0$ is evolved according to $\hat \rho(t)=\hat U(t) \hat \rho_0 \hat U^\dagger (t)$ with $\hat U(t)=\mathcal{T}\exp(-i\int_0^t\hat V_I(t')dt')$ . Up to second order in $\hat V$, the time evolution operator is given by
\begin{equation}
\hat U(t)= \mathbb{1}-i\int_0^t dt_1 \hat V_I(t_1)-\int_0^tdt_1\int_0^{t_1}dt_2 \hat V_I(t_1)\hat V_I(t_2)+\mathcal{O}(V^3).
\end{equation}
The expectation value of an observable $\hat A$ at time $t$, $\braket{\hat A(t)} \equiv \Tr\left(\hat \rho(t) \hat A_I(t) \right)$,  is then given by
\begin{align}
\braket{\hat A(t)} - \braket{\hat A(t)}_{V=0} &= -i \int_0^t dt_1 \braket{\left[\hat A_I(t), \hat V_I(t_1)\right]}+\int_0^tdt_1\int_0^{t}dt_2 \braket{\hat V_I(t_1) \hat A_I(t) \hat V_I(t_2)}\notag\\
&-\int_0^tdt_1\int_0^{t_1}dt_2 \braket{\hat V_I(t_2)\hat V_I(t_1) \hat A_I(t) + \hat A_I(t)\hat V_I(t_1)\hat V_I(t_2)}+\mathcal{O}(V^3),
\label{eq:response1}
\end{align}
where $\braket{\hat A(t)}_{V=0}$ is the expectation value in the absence of a probe pulse.

The above expression is valid for any density matrix $\hat \rho_0$, in particular also for thermal equilibrium $\hat \rho_0=\frac{1}{Z} e^{-\beta\hat H_0}$ with $Z=\Tr(e^{-\beta\hat H_0})$. Moreover, it offers a unified perspective on non-equilibrium dynamics between cold atom experiments, where usually $\hat \rho_0$ is thought off as some (product) initial state, given by the ground state of some Hamiltonian different to $\hat H_0$, i.e. a `` quantum quench'', and condensed matter pump-probe experiments, where $\hat \rho_0=\frac{1}{Z} e^{-\beta\hat H_0}$, but there is an additional strong, time dependent pump term during the time evolution $\hat H_0 \rightarrow \hat H_0+\hat H_\mathrm{pump}(t)$. In fact, these two perspectives are equivalent if the pump pulse only acts up to some time $t^*$ before the probe is turned on, as then the action of the pump pulse can be absorbed into the initial density matrix by $\hat \rho_0= \hat U_0 (t^*) \frac{1}{Z}e^{-\beta\hat H_0} \hat U_0^\dagger (t^*)$. 

In the following, we will specify this expression to the cases of the probe pulse coupling the atom creation/annihilation operator $c/c^\dagger$, which yields the cold-atom analogues of time-resolved ARPES and ARIPES, probing the two-time correlation functions $\braket{\hat c_\mathbf{q}^\dagger(t_1) \hat c_\mathbf{q}(t_2)}$, $\braket{ \hat c_\mathbf{q} (t_1)\hat c_\mathbf{q}^\dagger(t_2)}$ respectively, with momentum $\mathbf{q}$ and lattice site $j$.

\subsection{Time dependent ARPES}
The protocol consists in coupling the ``system'' optical lattice to another tube/layer  representing a ``detection'' lattice, which is initially empty. The detection lattice is offset by an energy $\Delta$, which is the analogue of the work function in condensed matter ARPES.
An analogue of a photopulse is created by modulating the lattice depth between system and detection layer, resulting in a coupling Hamiltonian
\begin{equation}
\hat V_I(t)= -t_y s(t) e^{-i\omega t}  \sum_\mathbf{k}\hat d^\dagger_{\mathbf{k}\sigma} \hat c_{\mathbf{k}\sigma}  + h.c. ,
\end{equation}
where $d^\dagger_{\mathbf{k}\sigma}$, $\hat c_{\mathbf{k}\sigma}$ creates/annihilates an atom with spin $\sigma$ in the detection/system layer; $t_y$ is the amplitude of the modulation, which needs to be small compared to the tunneling within the system; $\omega$ is the modulation frequency.  The detection system is assumed to be non-interacting, such that $\hat H_d=\sum_\mathbf{q} (\epsilon_\mathbf{q}+\Delta)\hat d^\dagger_{\mathbf{q}\sigma}\hat d_{\mathbf{q}\sigma}$ with $\epsilon_\mathbf{q}$ the non-interacting dispersion of the detection lattice. The operator measured in this scheme is the momentum space occupation number in the detection system $\hat A= \hat n_{\mathbf{q}\sigma}\equiv \hat d^\dagger_{\mathbf{q}\sigma} \hat d_{\mathbf{q}\sigma}$, which may be obtained from the band-mapping schemes in Ref.~\cite{Bohrdt2018}. The total initial density matrix $\hat \rho_0$ is a product state of the empty detection system and the system density matrix $\hat \rho_0=\ket{0}\bra{0}_d \otimes \hat \rho_s$.

Inserting $\hat{A}$ and $\hat V_I$ into Eq.~\ref{eq:response1}, we directly see that the linear term in $\hat V_I$ vanishes as it contains a vacuum expectation value of an odd number of detection system creation/annihilation operators. Furthermore, the terms in the second line of Eq.~\ref{eq:response1} also vanish as $\hat A_I(t)=\hat n_{\mathbf{q}\sigma}(t)$ acting on the empty detection initial state gives zero. Moreover, in the absence of system-detection layer tunneling, the occupation number in the detection system stays zero, such that $\braket{\hat n_{\mathbf{q}\sigma} (t) }_{t_y=0}=0$ at all times. The last remaining term then finally gives
\begin{align}
\braket{\hat n_{\mathbf{q}\sigma} (t) }&= t_y^2 \int_0^t dt_1\int_0^t dt_2 \,s(t_1)s(t_2) \sum_{\mathbf{k'}\mathbf{k''},\sigma'\sigma''} \braket{0|\hat d_{\mathbf{k'}\sigma'}(t_1)\hat n_{\mathbf{q}\sigma}(t)\hat d^\dagger_{\mathbf{k''}\sigma''}(t_2)|0} \Tr(\hat \rho_s \hat c^\dagger_{\mathbf{k'}\sigma'}(t_1)\hat c_{\mathbf{k''}\sigma''}(t_2)) e^{i\omega(t_1-t_2)}\notag\\
&= t_y^2 \int_0^t dt_1\int_0^t dt_2 s(t_1)s(t_2) e^{i(\omega-\epsilon_\mathbf{q}-\Delta)(t_1-t_2)} \Tr(\hat \rho_s \hat c^\dagger_{\mathbf{q}\sigma}(t_1)\hat c_{\mathbf{q}\sigma}(t_2))\\
&= t_y^2 \int_0^t dT \int_{-\tau_\mathrm{max}(T)}^{\tau_\mathrm{max}(T)} d\tau s\left(T-\frac{\tau}{2}\right)s\left(T+\frac{\tau}{2}\right) e^{i(\omega-\epsilon_\mathbf{q}-\Delta)\tau}A(T,\tau),
\end{align}
where we defined center of mass time $T=\frac{1}{2}(t_1+t_2)$, and relative time $\tau$ as well as the ``lesser'' Green's function $A(T,\tau)=\Tr(\hat \rho_s\hat c^\dagger_{\mathbf{q}\sigma}(T+\tau/2)\hat c_{\mathbf{q}\sigma}(T-\tau/2))$. The maximum relative time is $\tau_\mathrm{max}=2T$ for $T\leq t/2$ and $\tau_\mathrm{max}=2t$ for $T>t/2$. In most situations however, we can send $\tau_\mathrm{max}\rightarrow \infty$ due to the rapid decay of $A(T,\tau)$.
We note that this expression is valid for both Fermionic and Bosonic species because we assumed the detection system to be initially empty. 

In the following, we will discuss a few instructive limits of the above general expression. 

\subsubsection{Equilibrium limit}
We can recover the equilibrium result of Ref.~\cite{Bohrdt2018} by inserting $s(t)=1$ as well as using that in equilibrium, $\hat \rho_s=\frac{1}{Z} e^{-\beta \hat H_0}$, $A(T,\tau)$ only depends on $\tau$. The rate of tunneling to the detection system is then given by 
\begin{align}
\Gamma_{\mathbf{q}\sigma}(\omega)&= \frac{1}{t} \braket{\hat n_{\mathbf{q}\sigma} (t) }\\
&=t_y^2 \int_{-\infty}^\infty dt e^{i(\omega-\epsilon_\mathbf{q}-\Delta)t} \frac{1}{Z}\Tr(e^{-\beta \hat H_0} \hat c^\dagger_{\mathbf{q}\sigma}(t)\hat c_{\mathbf{q}\sigma})\\
&= t_y^2 A_{\mathbf{q}\sigma}(\omega-\epsilon_\mathbf{q}-\Delta).
\end{align}
$A_{\mathbf{q}\sigma}(\omega)$ is the hole spectral function
\begin{equation}
A_{\mathbf{q}\sigma}(\omega)=\frac{1}{Z}\sum_{nm}e^{-\beta E_n} |\braket{m|\hat c_{\mathbf{k}\sigma}|n}|^2 \delta(\omega-(E_m-E_n)).
\label{eq:specfunc_hole}
\end{equation}

\subsubsection{Gaussian pulse} 
A (normalized) Gaussian pulse centred around $t=t_p$ as $s(t)=\frac{1}{\sqrt{2\pi\sigma^2}}\exp(-(t-t_p)^2/2\sigma^2)$, leads to
\begin{equation}
\braket{\hat n_{\mathbf{q}\sigma} (t) }=\frac{1}{2\pi\sigma^2}\int_{0}^{t} dT \int_{-t_\mathrm{max}}^{t_\mathrm{max}} d\tau \exp\left(-\frac{(T-t_p)^2}{\sigma^2/2}\right)\exp\left(-\frac{\tau^2}{8\sigma^2}\right) e^{i(\omega-\epsilon_\mathbf{q}-\Delta)\tau} A(T,\tau),
\end{equation}
Typically, $A(T,\tau)$ decays rapidly as a function of $\tau$ (with the decay rate corresponding to the lifetime of excitations in the system), such that we can extend the integral boundaries for the $\tau$ integral to $\pm\infty$. Then we can interpret the $\tau$ integral as a Fourier transform, with a broadening introduced by the finite pulse length, leading to
\begin{equation}
\braket{\hat n_{\mathbf{q}\sigma} (t) }=\sqrt{\frac{2}{\pi\sigma^2}}\int_{0}^{t} dT  \exp\left(-\frac{(T-t_p)^2}{\sigma^2/2}\right)\int \frac{d \tilde \omega}{2\pi}\exp\left(-2\sigma^2\tilde\omega^2\right) A(T,\omega-\epsilon_\mathbf{q}-\Delta-\tilde \omega).
\end{equation}
Hence, a Gaussian pulse centred around $t_p$ measures the time dependent lesser Green's function averaged over a time and frequency window fulfilling the ``uncertainty relation'' $\sigma_T^2 \sigma^2_\omega = \frac{1}{4}$. 

\subsection{Time dependent ARIPES}
Here we show how to measure the time dependent particle spectral function in angle-resolve inverse photo-emission spectroscopy, in which atoms are injected rather than ejected from the system layer.

To measure the ARIPES spectrum, we propose to prepare the detection layer in a bandinsulator, i.e. all momenta are initially filled. Then, the same coupling term as in the ARPES protocol is turned on. In this case, the second line in Eq.~\ref{eq:response1} is not zero, however for Fermions it cancels with another contribution from the second term in the first line. In total, we get for Fermions
\begin{align}
1-\braket{\hat n_{\mathbf{q}\sigma} (t) }= t_y^2 \int_0^t dT \int_{-\tau_\mathrm{max}(T)}^{\tau_\mathrm{max}(T)} d\tau s\left(T-\frac{\tau}{2}\right)s\left(T+\frac{\tau}{2}\right) e^{-i(\omega-\epsilon_\mathbf{q}-\Delta)\tau}A^>_{\mathbf{q}\sigma}(T,\tau),
\end{align}
i.e. in this case the hole propagation in the detection layer needs to be measured. The result involves the particle spectral function
\begin{equation}
A^>_{\mathbf{q}\sigma}(t_1,t_2)=\Tr(\hat \rho_s \hat  c_{\mathbf{q}\sigma}(t_1) \hat c^\dagger_{\mathbf{q}\sigma}(t_2)),
\end{equation}
which measures the unoccupied states in the system.
The same manipulations as above go through analogously to obtain the equilibrium limits and the case of a Gaussian pulse. 
\subsection{Extracting $|A|$ from two copies of the same state}
Here we give an alternative protocol based on Ref.~\cite{Bohrdt2017}. It consists in evolving two copies of the system, acting with some operators on them and interfering them in the end.

First, prepare two copies of the same initial state $\ket{\Psi}\otimes\ket{\Psi}$ and let them evolve under the same Hamiltonian $\hat H$ for time $t_1$, such that we get $\exp{(-i\hat H t_1)}\ket{\Psi}\otimes\exp{(-i\hat H t_1)}\ket{\Psi}$. Then, remove a particle in the first copy at site $i$, evolve for time $(t_2-t_1)$ with $t_2>t_1$ and remove a particle in the second copy at site $j$. Thus, we end up in the state $\exp{(-i\hat H (t_2-t_1))}\hat c_i\exp{(-i\hat H t_1)}\ket{\Psi}\,\otimes\,\hat c_j\exp{(i\hat H t_2)}\ket{\Psi}$. Finally, measure the swap operator $ \mathrm{SWAP}$ by tunnel-coupling the two copies and measuring the parity-projected particle number, leading to 
\begin{align}
\braket{\mathrm{SWAP}}_{t_1<t_2}= |\braket{\hat c^\dagger_j(t_2)\hat c_i(t_1)}|^2,
\end{align}

which is the absolute value of the space and time dependent hole spectral function.

\end{document}